\documentclass[12pt]{article}  % Use the article class, 12pt font size
\usepackage{graphicx} % For including graphics
\usepackage[numbers]{natbib}
\usepackage{amsmath}    % For better math formatting
\usepackage{setspace}   % For line spacing control
\usepackage[margin=1in]{geometry}  % Adjust page margins
\usepackage{helvet}  % Use Helvetica (Arial-like font)
\usepackage{chemformula}  % Formula subscripts using \ch{}
\usepackage[T1]{fontenc}  % Use modern font encodings
\usepackage{authblk}      % For handling authors and affiliations more easily
\usepackage[version=4]{mhchem}

\title{Interaction of dopants with the I$_3$-type basal stacking fault in hexagonal-diamond Si}

\author[1]{Marc Túnica}
\author[1,2]{Perpetua Wanjiru Muchiri}
\author[1]{Alberto Zobelli}
\author[3]{Anna Marzegalli}
\author[3]{Emilio Scalise}
\author[1]{Michele Amato\thanks{michele.amato@universite-paris-saclay.fr}}

\affil[1]{Université Paris-Saclay, CNRS, Laboratoire de Physique des Solides, 91405 Orsay, France}
\affil[2]{School of Physics and Space Sciences, The Technical University of Kenya, P.O. Box 52428-00200, Nairobi, Kenya}
\affil[3]{Department of Materials Science, University of Milano-Bicocca, Via Roberto Cozzi 55, 20125 Milan, Italy}

\date{\today}

\begin{document}

\maketitle

%%%%%%%%%%%%%%%%%%%%%%%%%%%%%%%%%%%%%%%%%%%%%%%%%%%%%%%%%%%%%%%%%%%%%
%% The manuscript does not need to include \maketitle, which is
%% executed automatically.
%%%%%%%%%%%%%%%%%%%%%%%%%%%%%%%%%%%%%%%%%%%%%%%%%%%%%%%%%%%%%%%%%%%%%

% \begin{tocentry}

% Some journals require a graphical entry for the Table of Contents.
% This should be laid out ``print ready'' so that the sizing of the

% Inside the \texttt{tocentry} environment, the font used is Helvetica
% 8\,pt, as required by \emph{Journal of the American Chemical

% The surrounding frame is 9\,cm by 3.5\,cm, which is the maximum
% permitted for  \emph{Journal of the American Chemical Society}
% graphical table of content entries. The box will not resize if the
% content is too big: instead it will overflow the edge of the box.

% This box and the associated title will always be printed on a
% separate page at the end of the document.

% \end{tocentry}

\begin{abstract}
Recently synthesized hexagonal-diamond silicon, germanium, and silicon-germanium nanowires exhibit remarkable optical and electronic properties when compared to cubic-diamond polytypes. Because of the metastability of the hexagonal-diamond phase, I$_3$-type basal stacking faults are frequently observed in these materials. Understanding and modulating the interaction between these extended defects and dopants are essential for advancing the design and performance of these novel semiconductors. In the present study, we employ density functional theory calculations to investigate the interaction of extrinsic dopants (group III, IV, and V elements) with the I$_3$-type basal stacking fault in hexagonal-diamond silicon. Contrary to the behavior observed in cubic-diamond silicon with intrinsic stacking faults, we demonstrate that neutral and negatively charged $p$-type impurities exhibit a marked tendency to occupy lattice sites far from the I$_3$-type basal stacking fault. The interaction of acceptors with the planar defect reduces their energetic stability. However, this effect is much less pronounced for neutral or positively charged $n$-type dopants and isovalent impurities. The thermodynamic energy barrier to segregation for these dopants is small and may even become negative, indicating a tendency to segregate into the fault. Through a detailed analysis of structural modifications, ionization effects, and impurity-level charge density distribution, we show that the origin of this behavior can be attributed to variations in the impurity's steric effects and its wave function character. Finally, all these results are validated by considering the extreme case of an abrupt hexagonal/cubic silicon interface, where acceptor segregation from the cubic to the hexagonal region is demonstrated, confirming the behavior observed for $p$-type dopants near the I$_3$-type defect.
\end{abstract}

%%%%%%%%%%%%%%%%%%%%%%%%%%%%%%%%%%%%%%%%%%%%%%%%%%%%%%%%%%%%%%%%%%%%%


\section{\label{sec:introduction}Introduction\protect}
Silicon (Si), Germanium (Ge), and SiGe alloys are the building blocks of the microelectronics and nanoelectronics industry. However, the cubic-diamond (cub) structure in which they typically crystallize presents an indirect bandgap, limiting their use in optical devices. To overcome this limitation, several groups have recently succeeded in synthesizing group IV hexagonal-diamond (hex) polytypes in the form of nanowires (NWs)~\cite{GalvaoTizei2020, Vincent2014, Hauge2015, Hauge2017, Tang2017, Fadaly2020}. Indeed, while hex-Si remains an indirect bandgap material, both its indirect and direct bandgaps are reduced compared to cub-Si, making it promising for photovoltaic applications~\cite{Rodl2015, Keller2023}. Furthermore, Ge and SiGe alloys with a high concentration of Ge exhibit a direct bandgap~\cite{Fadaly2020, Wang2021, Borlido2021, Keller2023, Rodl2019, Amato2014, Xiang2022} which is
associated with a luminescence in the infrared spectral range.

Recent structural characterization studies of hex-Si, hex-Ge, and hex-SiGe have reported the presence of planar defects~\cite{Hauge2015, Fadaly2020, Fadaly2021, Vincent2022, Rovaris2024, Peeters2024}, with the I$_3$-type basal stacking fault (I$_3$-BSF) being the most common. This intrinsic defect results from the substitution of a B-plane by a C-plane in the typical ABABABAB stacking sequence of the hexagonal phase. Viewed along the $\left[0001\right]$ crystallographic direction, this defect manifests itself as a sequence of two twinned cubic cells with a hexagonal one in between, in other words the I$_3$-BSF introduces two symmetric rows of atoms with cubic-like bonding within the hexagonal structure~\cite{Fadaly2021}. Note that I$_3$-BSF usually does not extend through the nanowire, taking only a part of $\{$0001$\}$ planes. The defective planes terminate in $\langle1120\rangle$ directions with 30$^\circ$ partial dislocations, usually giving rise to a triangular-shaped defect~\cite{Vincent2022, Fadaly2021, Rovaris2024}. Neither the defective atomic layers nor the dislocations introduce any strain in the system. Indeed, the dislocations in the two adjacent defective planes have opposite  Burgers vectors, canceling the relative deformation field. Several studies have demonstrated that these defects may arise from the growth kinetics under specific conditions~\cite{Vincent2022, Rovaris2024}, and they do not introduce any additional states within the bandgap of hex-Si and hex-Ge~\cite{Fadaly2021}, thus not affecting the optical properties of the materials.

Nevertheless, planar defects can strongly influence dopant distribution and significantly impact the structural and electronic properties, as demonstrated in the case of cub-Si~\cite{Justo1999, Ohno2010, Ohno2012, Zhao2019, Yamamoto2014, Ohno2018}. Therefore, it is crucial to understand the impurity behavior close to extended defects, as dopant accumulation may affect their electrical activation and degrade device performance~\cite{InoueAPL2009,KuchiwakiSEMSC2008}. Although extrinsic doping in hex-Si, hex-Ge, and hex-SiGe has been experimentally achieved~\cite{Fadaly2020, Peeters2024}, characterizing the interaction of dopants with extended defects remains challenging due to the difficulty of providing a precise chemical characterization at the relevant defect scale. From a theoretical perspective, while doping and point defects in hex-Si and hex-Ge have already been studied by ab initio methods~\cite{Amato2019, Amato2020, Sun2021, Wang2021, TunicaPRM2024}, the effects of the I$_3$-BSFs on the doping properties remain unaddressed.

In this work, we employ density functional theory (DFT) simulations to investigate the interaction of $p$-type and $n$-type dopants with the I$_3$-BSF in hex-Si. Note that we consider I$_3$-BSF as infinitely extended (dislocations bounding the I$_3$-BSF are not modeled), looking at the local interaction of dopants with the defective layers. Through a careful structural and energetic analysis, we demonstrate that neutral and negatively charged acceptors prefer to occupy lattice sites far from the I$_3$-BSF. Nevertheless, this behavior is less marked for ionized acceptors. Neutral and positively charged donors and isovalent impurities do not show the same trend because their segregation energy is small and can be negative when the dopant is close to the fault.
A detailed analysis of structural modifications, ionization effects, and charge density distributions proves that the origin of this distinct behavior can be ascribed to a variation in the steric effect and the dopant wave function localization.

We demonstrate that the segregation of $p$-type dopants towards a hexagonal-like environment is also present in the extreme case of an abrupt hex/cub Si interface, thus confirming the results obtained for $p$-type dopants in the proximity of the I$_3$-BSF. Our findings offer inspiration for hex-Si materials processing, such as identifying conditions that influence defect formation and dopant incorporation and suggesting novel opportunities, such as exploiting anisotropic segregation tendencies to build hex/cub-Si heterostructures with functional doping profiles.

\section{\label{sec:methodology}Computational Details\protect}
Spin-polarized ab initio DFT simulations were performed using the SIESTA code~\cite{Soler2002}. The generalized gradient approximation (GGA) with the Perdew, Burke, and Ernzerhof (PBE) functional~\cite{Perdew1996} was used. Since experimentally grown hexagonal nanowires typically have diameters greater than 10 nm, quantum confinement effects on the bandgap can be neglected ~\cite{Amato2010, Amato2012, Mohammad2014, Rurali2010, GalvaoTizei2020, DavidNL2017}. Therefore, we used bulk supercells as an accurate and reliable approximation for these larger-diameter nanowires. A double-$\zeta$ polarized basis set was used for the valence electrons of Si, while a double-$\zeta$ basis set plus two polarization orbitals was employed for the dopant atoms. Inner core electrons were replaced by norm-conserving pseudopotentials of the Troullier-Martins type~\cite{Troullier1991}. The hex-Si crystal cell was optimized using an 8$\times$8$\times$8 k-point grid generated with the Monkhorst-Pack sampling~\cite{Monkhorst1976}. The optimized lattice parameters, $a_{Si} =3.84$~{\AA} and $c_{Si}=6.34$~{\AA}, are in good agreement with values reported in previous works employing the same type of exchange-correlation functional~\cite{Keller2023, Wang2021}. Extrinsic doping was analyzed in the hex-Si crystal, both pristine and with an I$_3$-type BSF. A 4$\times$4$\times$6 supercell (see Figure~\ref{fig:geometry_of_the_simulation_sf}a) was used. These cell dimensions are sufficiently large to minimize the spurious electrostatic interaction between the periodic images of dopants and BSFs. The structure is composed of ten layers of hex-Si along the $\left[0001\right]$ direction (c-axis of the hexagonal structure), interspersed with two cubic-diamond stoichiometry planar defects (see the light blue rectangles in Figure~\ref{fig:geometry_of_the_simulation_sf}a) and a hexagonal region between them. The supercell contains 384 atoms (corresponding to a volume of 7.8~$\cdot$~$10^{-21}~\text{cm}^{3}$) and, when doped, leads to a doping concentration of 1.3~$\cdot$~$10^{20}~\text{cm}^{-3}$ which is one or two orders of magnitude larger than the one measured in the 2H experimental samples~\cite{Fadaly2020, Peeters2024} and consistent with values observed in high-doping regimes for 3C-Si and 3C-SiGe nanowires~\cite{NahAPL2008,NahAPL2009,WangNL2005}. This concentration has not only been achieved experimentally but has also been previously explored in theoretical studies~\cite{AmatoJCE2012,AmatoJAP2012} providing a relevant basis for our computational framework. Moreover, since specific non-linear effects are typically observed for concentrations above $10^{20}~\text{cm}^{-3}$ (see, for example, Refs.~\cite{XiePRL1999, SolmiJAP1998, LarsenJAP1993}), the trends discussed here in dopant stability such as dopant-induced strain, site preference, and formation energies are expected to remain consistent across the range of doping levels considered.

We considered common substitutional dopant atoms in Si: acceptors (B, Al, Ga, In), donors (N, P, As, Sb), and isovalent impurities (C, Ge). All the atomic positions in doped and undoped systems were optimized using a conjugate gradient algorithm that calculates DFT forces through the Hellmann-Feynman theorem. Structural convergence was achieved when the forces were less than 10$^{-2}$~eV/Å and the stress was lower than 10$^{-1}$~GPa. A 2$\times$2$\times$1 Monkhorst-Pack k-point grid and a mesh cutoff of 500~Ry were carefully chosen to ensure the correct convergence of total energies and forces. For charged impurity calculations, an electron was added or subtracted with the inclusion of a compensating jellium background to maintain charge neutrality and avoid divergence of the electrostatic energy. The structures containing the charged impurities were optimized, starting from the previously relaxed neutral-doped configurations, using the same convergence parameters setup.

The results obtained for hex-Si with the I$_3$-BSF were then compared with the case of an abrupt hex/cub-Si interface (Figure~\ref{fig:geometry_of_the_simulation_sf}b), built along the $\left[0001\right]$ hexagonal crystal direction. Such an interface comprises eight hexagonal and six cubic layers and has 448 atoms. Even in this case, the doping concentration is about $10^{20}~\text{cm}^{-3}$. By calculating DFT ground-state total energy and the corresponding charge density distribution, we investigated the effect of the I$_3$-BSF on the dopant behavior and the electronic properties of the system. To this end, we relaxed several defect configurations with a substitution atom positioned at different layers along the c-axis (highlighted in red in Figure~\ref{fig:geometry_of_the_simulation_sf}), and for each configuration, we calculated the dopant segregation energy (DSE). The relaxed structures of the studied doped systems are attached in the SI. As in previous works~\cite{Ohno2018,Ohno2012}, the DSE is defined here as the difference between the ground state energy of the doped system with the fault, $E_\text{tot}^{I3-BSF} (z)$, and reference energy where the interaction with the stacking fault is minimized. The supercell size was hence converged to ensure that the total energy of the system with the dopant positioned at the farthest distance from the I$_3$-BSF (denoted as $z_0$) coincides with the sum of the doped pristine supercell and the stacking fault energy (the difference between the two values is less than 0.01 eV). Therefore, the DSE is expressed as

\begin{equation}\label{eq:segregation_energy}
DSE (z) = E_\text{tot}^{I3-BSF} (z) - E_\text{tot}^{I3-BSF} (z_0),
\end{equation}
where $E_\text{tot}^{I3-BSF}(z)$ is the total energy of the defective system with the dopant at position $z$, and $E_\text{tot}^{I3-BSF}(z_0)$ is the total energy when the dopant is at the farthest atomic plane from the defect. Negative values of DSE correspond to an attraction between the I$_3$-BSF and the dopant, while positive values indicate that segregation is thermodynamically unfavored. The DSE provides a quantitative evaluation of the dopant-stacking fault interaction and is determined by two primary contributions: the structural relaxation energy of the lattice induced by the dopant and the electrostatic interaction between the dopant and the I$_3$-BSF.

\section{\label{sec:Results}Results and discussion\protect}
We start our analysis by calculating the stacking fault energy ($\gamma^{I3-BSF}$) defined as the energy difference normalized by the surface area between the total ground-state energy of the supercell containing the stacking fault, $E_{tot}^{I3-BSF}$, and that of the pristine hexagonal phase, $E_{tot}^{Pristine}$. This can be expressed as:
\begin{equation}\label{stacking_fault_energy}
\gamma^{I3-BSF} = \dfrac{E_{tot}^{I3-BSF} - E_{tot}^{Pristine}}{A},
\end{equation}
We found $\gamma_{Si}^{I3-BSF}\approx -85~\text{mJ/m}^2$, which is compatible with the metastability of the hexagonal phase, whose total DFT ground-state energy is nearly 10-20 meV higher than the Si cubic phase~\cite{Raffy2002}. These findings also agree with the negative formation energies reported in the literature~\cite{Fadaly2021} suggesting that defects like I$_3$-type BSFs are energetically stable in such crystals. The thermodynamic preference may be related to the distinct symmetries of hexagonal and cubic polytypes, associated with different total energies. In particular, the Si hexagonal pristine structure exhibits a C$_{3v}$ symmetry, characterized by three equal bonds in the $[0001]$ basal plane and a longer one perpendicular to it, which leads to a reduced bond strength along this direction.
On the other hand, cubic crystal structures, such as those introduced by the I$_3$-BSF, display a perfect tetrahedral symmetry ($T_d$), where four equivalent bonds match the $sp^3$ hybridization of the four valence electrons in Si, thus inducing higher chemical stability and lower total energy if compared to hex-Si. Moreover, it is worth noting that previous works~\cite{Amato2019, Amato2020,TunicaPRM2024} demonstrated that the stability of a dopant in one of the two bulk fault-free phases also depends on a complex interplay between the local structural environment around the dopant (induced by the symmetry of the host crystal), its atomic radius as well as and its valence. In particular, in hex-Si, acceptors have lower formation energy, present a $C_{3v}$ symmetry, and have charge transition levels shallower with respect to the cub-Si case. On the other hand, donors are more stable in the cubic phase where they adopt a $T_d$ symmetry and have transition levels energies shallower with respect to hex-Si~\cite{Amato2019, Amato2020,TunicaPRM2024}.

\begin{figure}
    \centering
    \includegraphics[width=\columnwidth,height=0.7\textheight,keepaspectratio]{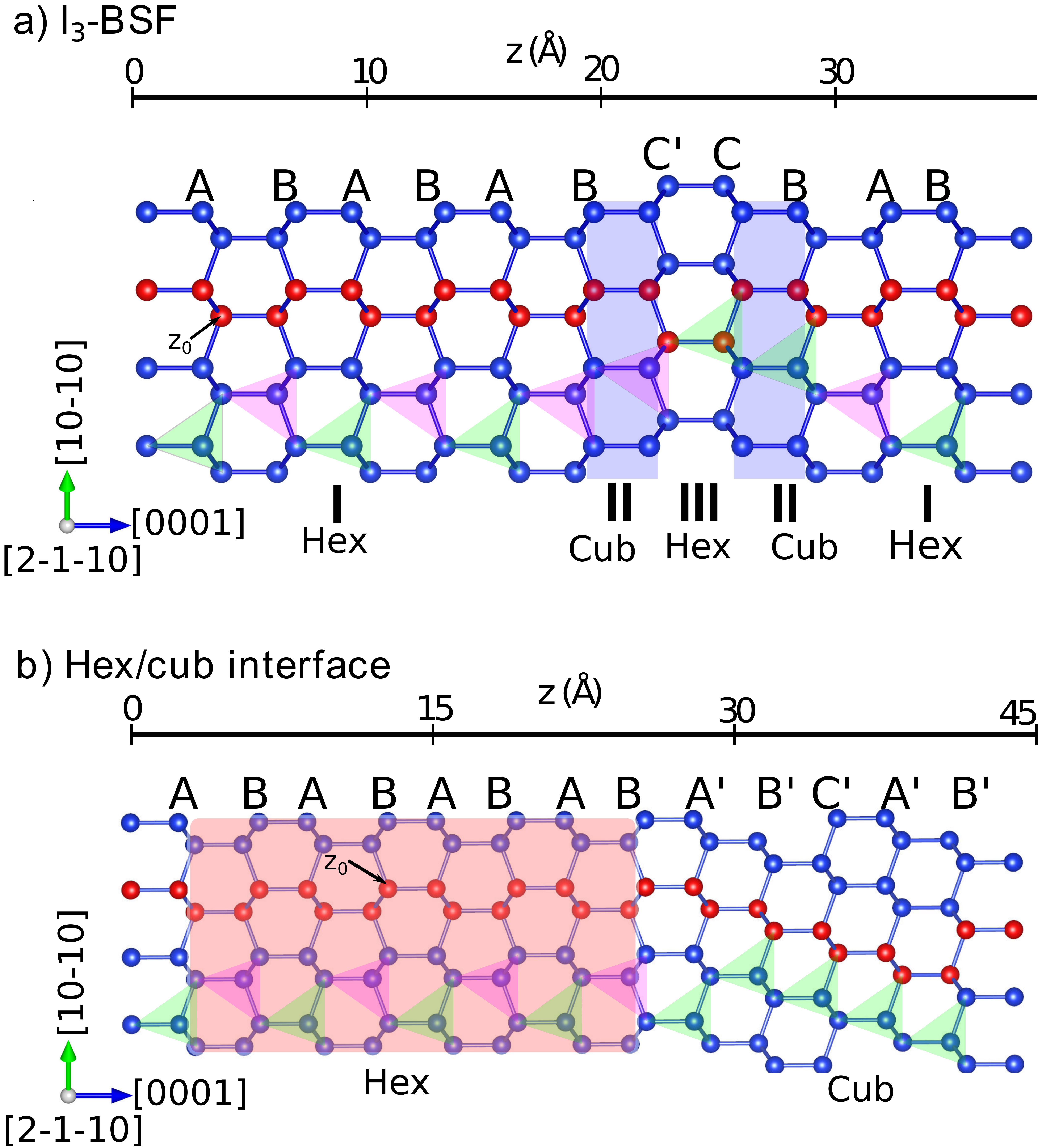}
    \caption{a) Supercell of the hex-Si phase crystal containing an I$_3$-BSF. Labels under the structure (I, II, and III) correspond to three distinct crystallographic arrangements within the material. Region I: pristine hexagonal structure. Regions II (blue rectangles): two cubic arrangements induced by the I$_3$-BSF. Region III: a hex-type structure between the two cubic regions.
    b) Hexagonal/cubic interface. The red rectangle corresponds to the atoms with a hexagonal symmetry, and the other side of the interface presents a cubic symmetry type. For both plots, red atoms correspond to the to the position of the dopants along the c-axis direction. $z_{0}$ corresponds to the farthest atomic plane from the planar defect and the reference of the energy on Eq.~\ref{eq:segregation_energy}. The pink and green triangles highlight the twinned or normal tetrahedra~\cite{ScalisePRA2019,HongpPMA2000,Marzegalli2024}. To further clarify the local symmetry across different regions, the stacking sequence (e.g., ABAB, ABCABC, etc.) is indicated in each panel.}
    \label{fig:geometry_of_the_simulation_sf}
\end{figure}

The hexagonal structure containing the I$_3$-BSF, shown in Figure~\ref{fig:geometry_of_the_simulation_sf}a, can be conceptually divided into three regions. Region I corresponds to the hexagonal Si bulk-like crystal structure (ABABAB in Figure~\ref{fig:geometry_of_the_simulation_sf}a). Region II includes the two cubic arrangements induced by the I$_3$-BSF, which breaks the hexagonal stacking sequence (BC' and BC in Figure~\ref{fig:geometry_of_the_simulation_sf}a). Region III lies between these two cubic sections and presents a hex-type structure (C'C in Figure~\ref{fig:geometry_of_the_simulation_sf}a).

\begin{figure*}
    \centering
    \includegraphics[width=0.32\textwidth]{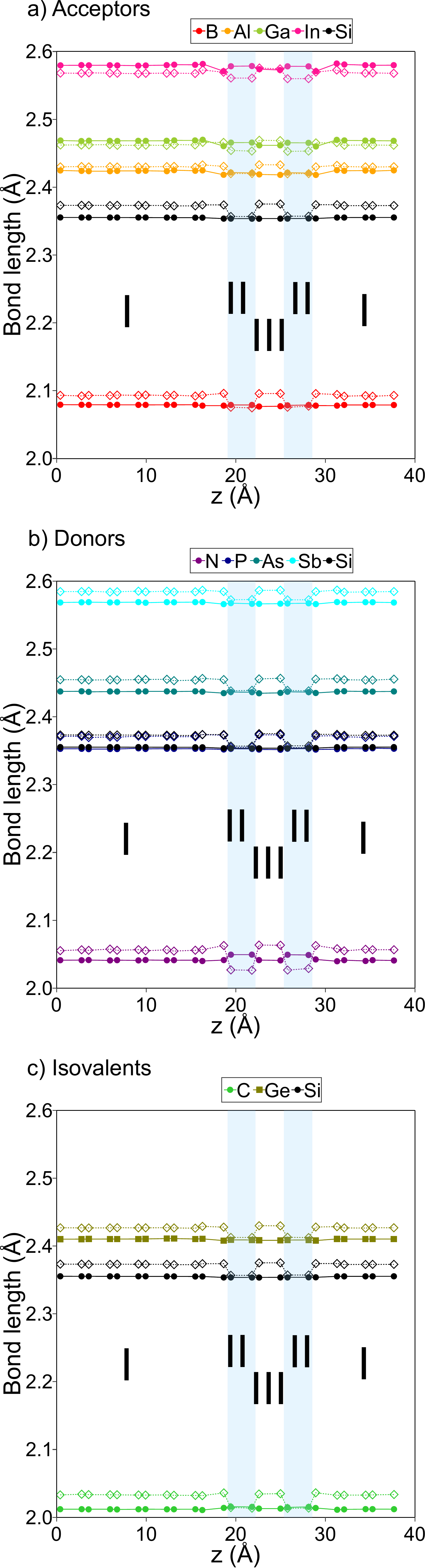}
    \caption{Neutral dopant first-neighbors bond distances in defective hexagonal-diamond Si: (a) acceptors, (b) donors, and (c) isovalent dopants bond lengths along the $[0001]$ direction. Blue regions correspond to the rows of atoms with the cubic-diamond arrangement introduced by the I$_3$-BSF (Regions II of Fig.~\ref{fig:geometry_of_the_simulation_sf}). Continuous lines with solid circles correspond to the averaged bond distances in the basal plane, and dashed lines with empty diamonds correspond to bond distances along the [0001] direction.}
    \label{fig:bonds_sf_si}
    \vspace{-9pt}
\end{figure*}

We begin the structural analysis by examining the local symmetry around the dopant. The dopant first-neighbor bond lengths in the $\{0001\}$ basal plane were averaged and compared to those along the perpendicular axis. As previously discussed,  both donors and acceptors preserve the $T_d$ symmetry in bulk cub-Si where all the dopant's first-neighbor distances are equivalent~\cite{CantelePRB2005,HakalaPRB2000,ZhouPRB2005}. This situation changes in the case of hex-Si, where both acceptors and donors are characterized by a marked C$_{3v}$ symmetry~\cite{TunicaPRM2024}.
Figure~\ref{fig:bonds_sf_si} shows the averaged bond distances of the dopant first-neighbor in the basal plane (continuous lines with solid circles) and along the c-axis direction (dashed lines with empty diamonds) for i) acceptors, ii) donors, and iii) isovalent impurities in hex-Si, located at the positions highlighted in red in Figure~\ref{fig:geometry_of_the_simulation_sf}a. When the two lines overlap, the local symmetry is $T_{d}$ while it becomes C$_{3v}$ when they differ. Each subplot includes reference data from the undoped structure for comparison (black lines). For the undoped hex-Si system containing the I$_3$-BSF, there is a clear transformation in the local symmetry, which transitions from C$_{3v}$ in Region I to $T_d$ in Region II and back to C$_{3v}$ in Region III.
When a dopant is introduced, the bond distances expand or contract depending on its covalent radius. As shown in Figure~\ref{fig:bonds_sf_si}a and demonstrated in other works~\cite{Amato2020,CantelePRB2005,HakalaPRB2000,ZhouPRB2005}, shorter interatomic distances in bulk Si correlate with smaller dopant covalent radii. However, from Figure~\ref{fig:bonds_sf_si}a, it is clear that the distance of the dopant from the I$_3$-BSF influences its bond lengths and local symmetry.

In particular, in Figure~\ref{fig:bonds_sf_si}a, a transition from C$_{3v}$ to $T_{d}$ in Region II (BC' and CB layers) is clear for B and Al. The same symmetry transformation holds for larger radius acceptors at the interfaces between Region I and II and between Region II and III. These acceptors also show a C$_{3v}$ symmetry in the cubic Region II. Interestingly, the bond lengths along the $\left[0001\right]$ direction are shorter for these cases than those in the basal plane. For donors (Figure~\ref{fig:bonds_sf_si}b) and for isovalent dopants (Figure~\ref{fig:bonds_sf_si}c), the symmetry around impurities also passes from C$_{3v}$ to $T_{d}$ in Region II. A noteworthy exception to this behavior is represented by the case of N (Figure~\ref{fig:bonds_sf_si}b). Indeed, although most of the dopants considered in cub-Si maintain a tetrahedral symmetry, both theoretical and experimental works demonstrated that N presents a local trigonal symmetry due to its low solubility in cub-Si~\cite{Amato2019,Amato2020,Brower1982,TaguchiJAP2005}.

Looking at Figure~\ref{fig:bonds_sf_si}a, it is clear that the effect of the I$_3$-BSF on the dopant local symmetry in hex-Si is negligible when the dopant is far from the planar defect. The impact is merely localized to the lattice sites in proximity of the planar defect, with bond distances and symmetries in the various crystallographic regions matching those observed in doped bulk materials~\cite{Amato2019, Amato2020,TunicaPRM2024}. When the impurity lies in the bulk-like region far from the stacking fault, for both $n$-type and $p$-type dopants, the local symmetry is C$_{3v}$ as demonstrated in the case of doped bulk hex-Si~\cite{Amato2019, Amato2020,TunicaPRM2024}. If the dopant occupies positions close to the I$_3$-BSF, there is a transition from the C$_{3v}$
to the $T_{d}$, driven by the presence of the two symmetric cubic layers in the I$_3$-BSF.

\subsection{Segregation energy\protect}
\begin{figure*}
    \centering
    \includegraphics[width=0.35\textwidth]{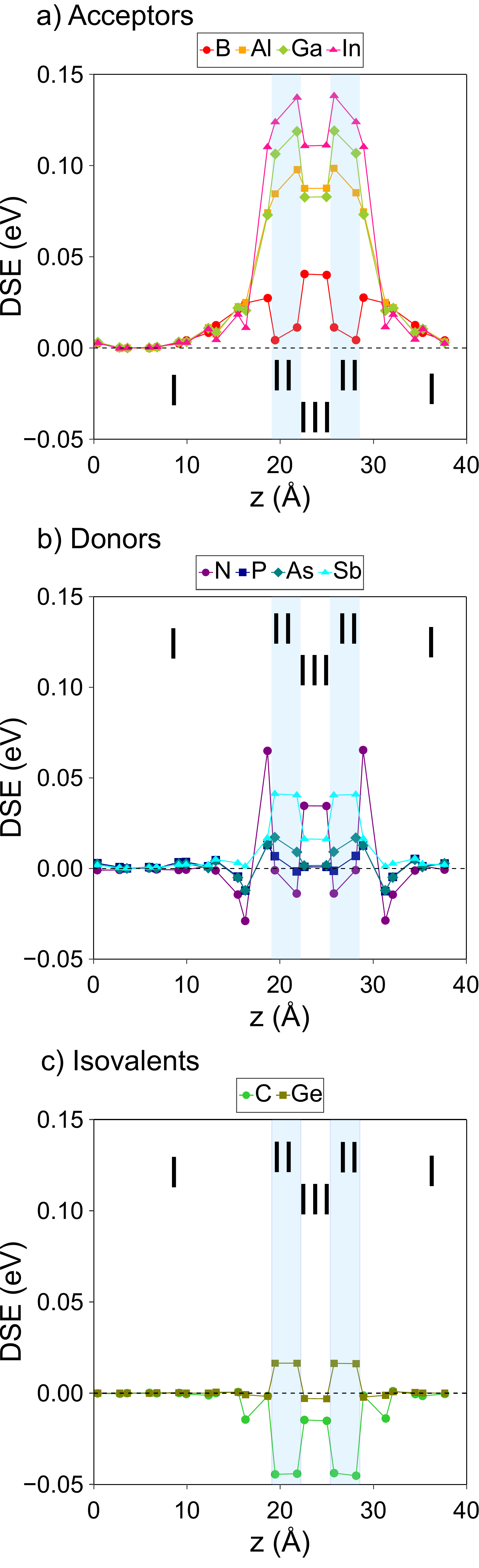}
    \caption{Segregation energy of neutral dopants in defective hexagonal-diamond Si: (a) acceptors, (b) donors, and (c) isovalent dopants along the $[0001]$ direction. Blue regions correspond to the rows of atoms with the cubic-diamond arrangement introduced by the I$_3$-BSF (Regions II of Fig.~\ref{fig:geometry_of_the_simulation_sf}).}
    \label{fig:energy_sf_si}
\end{figure*}

Figure~\ref{fig:energy_sf_si} shows the dopant segregation energy (DSE) in hex-Si, calculated using Eq.~\ref{eq:segregation_energy}, for acceptors, donors, and isovalent dopants along the red path of Figure~\ref{fig:geometry_of_the_simulation_sf}a. As shown in other works~\cite{ZiebarthPRB22015,ZiebarthPRB2015}, a negative value of this quantity suggests that the attraction between the dopant and the planar defect is favored. In contrast, a positive value avoids the segregation toward the stacking fault. For all the neutral acceptors~\textemdash~with the noticeable exception of B~\textemdash, the DSE is always positive close to the I$_3$-BSF, reaching maximum values of 80-140~meV in Region II and 70-110~meV in Region III. The significant DSE differences between Regions I (hexagonal bulk-like environment) and II/III (basal stacking fault) suggest a thermodynamic driving force for acceptors to occupy lattice sites in the hexagonal pristine region and a dopant depletion near the I$_3$-BSF. These findings may be put in relation to the work of Yamamoto et al.~\cite{Yamamoto2014}, in which it was theoretically predicted that dopants in cub-Si tend to segregate towards single intrinsic stacking faults. The negative segregation energies they calculated (100~meV for Al, 90~meV for Ga, and 130~meV for In) are of the same magnitude as the thermodynamic energetic barriers obtained here but present an opposite sign. Our results show that B does not follow this trend, presenting a very low DSE of approximately 40~meV in Region III and 30~meV at the boundary between Regions I and II.
This particular result was further confirmed by plane waves DFT-GGA simulations we performed using the Quantum Espresso code~\cite{Giannozzi2009}. Though unexpected, this behavior for B was also observed in the case of cub-Si containing intrinsic stacking faults~\cite{Yamamoto2014}, and it is related to the significant radius mismatch between B and Si atoms.

As for neutral donors (Figure~\ref{fig:energy_sf_si}b), DSEs are less pronounced than for acceptors, but neither here any relevant segregation into the stacking fault are predicted. The maximum DSEs in Region II are around 40~meV for Sb and 20~meV for As and P. These DSEs are smaller than the ones reported in the case of cub-Si containing an intrinsic stacking fault~\cite{Justo1999, Yamamoto2014}. In the case of N, our results show a maximum energy peak of 70~meV between Region I and II and around 35~meV in Region III. This difference in the trend of DSE can be associated with the N particular symmetry in Si and its minimal atomic radius.
Finally, the DSEs of C and Ge in Region II are lower than 20~meV, being even negative for C, suggesting a possible segregation behavior for this impurity.

As clearly explained in Ref.~\cite{AntonelliJAP2002}, the DSE may be separated into two main contributions: the elastic energy that is due to the packing local rearrangement of the dopant and the long-range electrostatic interaction of the dopant with the stacking fault. Though it is complicated to extract and separate the two contributions from the calculated DFT total energy, we tried to evaluate them by looking at the dependence of the DSE on the dopant charge state and the covalent radius mismatch between the dopant and the Si host atom.

\begin{figure*}
    \centering
    \includegraphics[width=\textwidth]{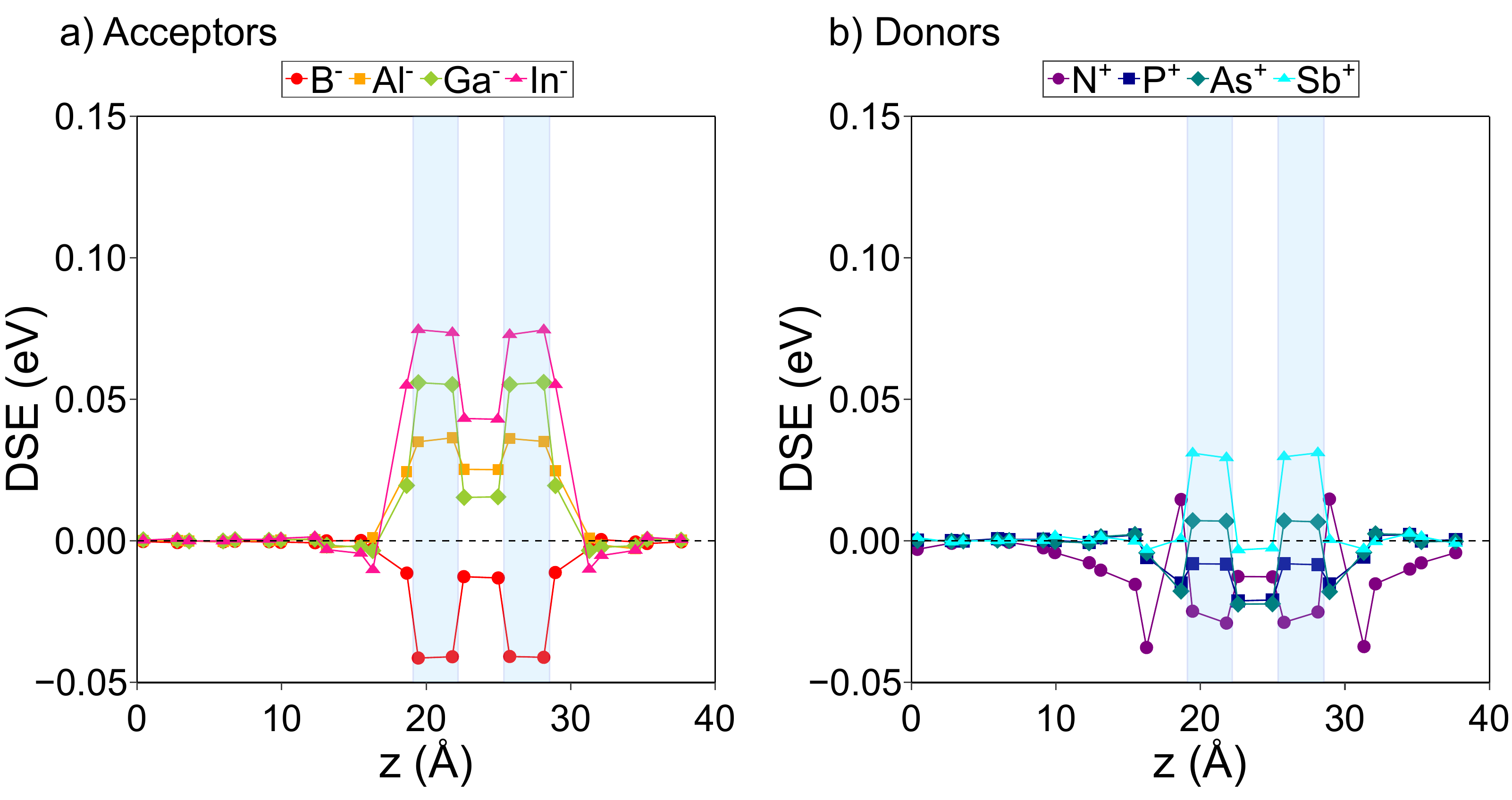}
    \caption{Segregation energy of ionized dopants in defective hexagonal-diamond Si: (a) negative acceptors, (b) positive donors along the $[0001]$ direction. Blue regions correspond to the rows of atoms with the cubic-diamond arrangement introduced by the I$_3$-BSF (Regions II of Fig.~\ref{fig:geometry_of_the_simulation_sf}).}
    \label{fig:energy_sf_si_charged}
\end{figure*}

\paragraph{Ionization effects.}To examine the role of the dopant charge state, we evaluate the DSE in the case of negatively charged acceptors and positively charged donors here reported in Figure~\ref{fig:energy_sf_si_charged}a and Figure~\ref{fig:energy_sf_si_charged}b. Extended supercells have been used (see Figure~\ref{fig:geometry_of_the_simulation_sf}a) to minimize the spurious long-range interactions between periodic images~\cite{Komsa2012}. The residual electrostatic contribution for a given dopant is independent of its position. Therefore, any other spurious Coulomb term are largely canceled out since the DSE is calculated as a difference in energy between different dopant configurations (see Eq.~\ref{eq:segregation_energy}).

Looking at Figure~\ref{fig:energy_sf_si_charged}, one can observe that when an electron is added or removed, the DSE in the proximity of the I$_3$-BSF is reduced. In particular, the DSE reduction  for Al$^{-}$, Ga$^{-}$, and In$^{-}$ is 43-62~\% in Region II, and 61-80~\% in Region III. Despite this decrease, DSEs for In$^{-}$, Ga$^{-}$ are still relevant, with values of 73~meV and 55~meV in Region II. Al$^{-}$ presents a less significant DSE of 34~meV in Region II. On the other hand, B$^{-}$ has a negative segregation energy of -40~meV close to the fault, indicating higher stability in the cubic-like region.
Regarding positively charged donors, the DSE is strongly reduced to negligible or negative values in Region II. These results allow us to depict a first general trend: neutral and charged $p$-type dopants prefer to occupy lattice sites far from the stacking fault plane, while neutral and charged $n$-type impurities do not have any clear preference. The ionization reduces the thermodynamic energy barrier to segregation with respect to the neutral case. This means charged dopant ions reveal a reduced interaction with the I$_3$-BSF. However, as the DSE has an electrostatic and elastic component, the ionization effect alone cannot explain the difference in the DSE trend among $p$-type and $n$-type dopants.

\begin{figure*}
    \centering
    \includegraphics[width=\textwidth]{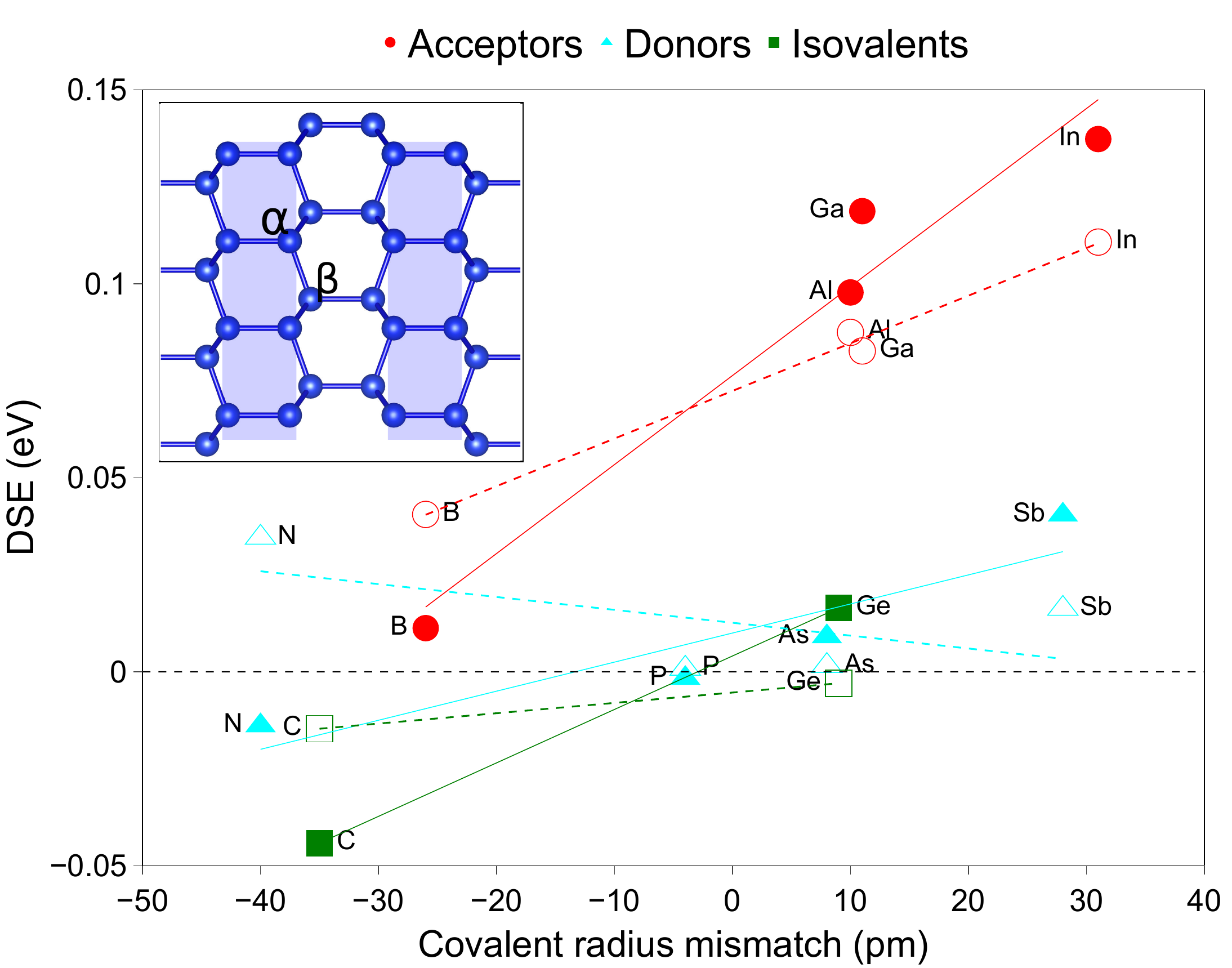}
    \caption{Neutral dopant segregation energy as a function of the covalent radius mismatch with Si for acceptors (red circles), donors (cyan triangles), and isovalent impurities (green squares) in the two lattice sites of Region II (filled markers, position $\alpha$) and III (empty markers, position $\beta$) of Figure~\ref{fig:geometry_of_the_simulation_sf}, as shown in the inset (left corner). Blue regions in the inset correspond to the rows of atoms with a cubic-diamond arrangement introduced by the I$_3$-BSF (Regions II of Fig.~\ref{fig:geometry_of_the_simulation_sf}). To guide the eye, linear regressions are shown for acceptors, donors, and isovalent defects. Solid lines indicate regressions fitted to filled data points; dashed lines correspond to those fitted to unfilled points.}
    \label{fig:energy_mismatch_si}
\end{figure*}

\paragraph{Steric effects.} To give an estimation of the elastic energy involved in the dopant-stacking fault interaction, we report in Figure~\ref{fig:energy_mismatch_si} the magnitude of the DSE as a function of the covalent radius mismatch with the Si atom for all the considered neutral dopants. As can be seen, the DSE increases almost linearly with the radius mismatch. In particular, dopants with larger radius compared to Si (In, Ga, Al) have positive and larger values of the DSE. When the radius mismatch gets smaller (as in the case of Ge, P, and As), the DSE is very close to zero. Finally, very small radius dopants (N, B, and C) induce a zero or negative DSE. As shown in the linear regressions reported in Figure~\ref{fig:energy_mismatch_si}, the steric effect seems to be more pronounced for acceptors than for donors or isovalents. Further confirmation of these findings is given by comparing the DSE for all the considered cases with and without geometrical optimization, as reported in Figure S1 of the Supporting Information (SI). The difference in DSE between relaxed and unrelaxed systems arises from the structural modifications occurring when the system is allowed to relax. In the unrelaxed system, the dopant-induced strain is not taken into account, whereas the relaxed system includes the energetic effects of atomic repositioning and strain relief, which are crucial for describing the dopant's stability. As observed in Figure S1, when the lattice mismatch is large, the difference between the relaxed and unrelaxed DSE is significant (for B, Ga, In, Al), while it is negligible when the radius mismatch is small (for P, As, Ge). This behavior is due to the strain induced by the dopant in the host lattice and is consistent with the results reported in Figure~\ref{fig:energy_mismatch_si}. For large positive (negative) lattice mismatches, the unrelaxed configuration does not account for the strain, leading to an underestimation (overestimation) of the DSE. Once the system is relaxed, atomic adjustments accommodate the dopant, and the DSE increases to reflect the energetic cost of this strain and lattice distortion.

More in general, these results complement the physical picture drawn with the ionized impurity results: the I$_3$-BSF defect represents a thermodynamic barrier for acceptors because of both an electrostatic Coulomb repulsion and the large positive radius mismatch with Si. In contrast, donors — except for N — do not see the I$_3$-BSF as a barrier, in particular when they are ionized. In this case, the elastic energy variation due to the covalent radius mismatch is not relevant to the interaction with the fault. As for isovalent impurities, the DSE is negligible for Ge, as expected, while it is negative for C due to the significant negative radius mismatch. This means that the I$_3$-BSF could represent an energy sink for C atoms.
Finally, it is worth noting that B represents a particular exception to the discussed trends. Indeed, ionization effects and radius mismatch are much more pronounced for this dopant than other acceptors. As in the case of C, B$^-$ atoms could segregate into the staking fault with negative DSE values.

\begin{figure*}
    \centering
    \includegraphics[width=0.8\textwidth]{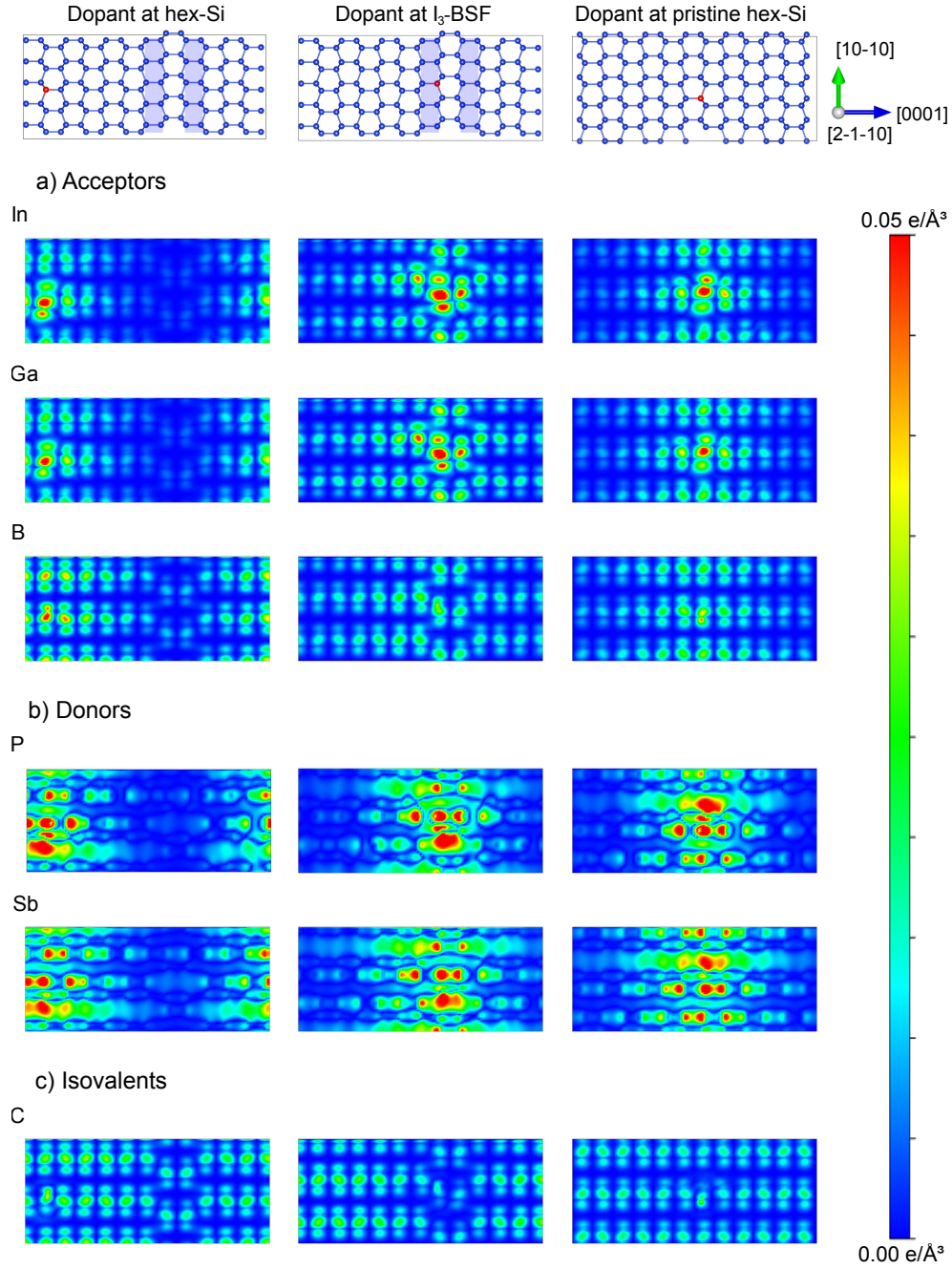}
    \caption{Charge density distribution maps of the impurity energy level at the $\Gamma$ point for (a) acceptors (In, Ga, B), (b) donors (P, Sb), and (c) isovalents dopants (C) when the dopant occupies a lattice site far from the fault (left column), close to the I$_3$-BSF (central column) and in the pristine hex-Si structure (right column). The projection is made on the $\{$2-1-10$\}$ plane crossing the dopant atom. Charge densities are calculated with the denchar post-processing program within the SIESTA package~\cite{Soler2002} and visualized using the VESTA software~\cite{MommaJAC2011} under a permissive acknowledgment-based license.}
    \label{fig:charge_density}
\end{figure*}

\paragraph{Symmetry effects.} As previously pointed out, for both $n$-type and $p$-type dopants, the interaction with the I$_3$-BSF is associated with a change in the local symmetry. In particular, the $C_{3v}$ symmetry observed in pristine hex-Si and lattice sites far from the fault becomes $T_{d}$ due to the presence of the two cubic layers (see Figure~\ref{fig:geometry_of_the_simulation_sf} and Figure~\ref{fig:bonds_sf_si}). To understand if this change could affect the symmetry and localization of the dopant wave function and the DSE trends, we carefully analyzed the charge density distribution for all the considered neutral dopants, moving from the hexagonal bulk-like environment toward the stacking faults planes.
We first analyzed the spatial distribution of the charge density in the I$_3$-BSF system without any impurity (see Figure S2 of the SI). As reported in previous theoretical works~\cite{Fadaly2021}, the presence of the  I$_3$-BSF does not perturb too much the valence band maximum (VBM) and conduction band minimum (CBM) if compared with the pristine hex-Si. However, the VBM charge density is depleted for the defective structure at the stacking fault region. This means, as already observed in the case of wurtzite GaN with cubic stacking faults~\cite{SchmidtPRB2002}, that the I$_3$-BSF has a long-range influence of the VBM wave function (see Figure S2 of the SI).
The behavior is different when an impurity is added to the system. In Figure~\ref{fig:charge_density}, the charge density distribution associated with the defect level at the $\Gamma$ point is plotted for acceptors, donors, and isovalent impurities in three different cases: i) when the dopant is far from the stacking fault (left column), ii) when the dopant occupies a lattice site close to the stacking fault plane (central column), and iii) when the impurity atom is added to the pristine hex-Si bulk structure (right column).
As it can be evinced from Figure~\ref{fig:charge_density}a, in the case of acceptors (In, Ga, B), the impurity level wave function has a marked $C_{3v}$ nature in the pristine hex-Si and the lattice sites away from the fault plane. Indeed, in both cases (left and right columns of Figure~\ref{fig:charge_density}a), the electron density at the two bonds of the dopant atom lying in the plane of the cut is different, with the bond normal to the stacking fault having a lower density than the other one. This is due to the trivalent valence of acceptors whose ground state wave function fits the $C_{3v}$ symmetry of the host hex-Si. In contrast, when the acceptor is near the stacking fault, the dopant wave function is more spatially localized, assuming a clear $T_{d}$ character due to the previously described structural changes. Indeed, in this case (central column of Figure~\ref{fig:charge_density}a), the electron density at the two bonds of the dopant atom lying in the plane of the cut is very similar. This change in the symmetry is corroborated by the Mulliken population analysis~\cite{MullikenJCP1955} presented in the SI. This further contributes to the explanation of the segregation trends: the DSEs close to the I$_3$-BSF plane are significant and positive because of the strong steric effect that forces the impurity level wave function to lose its $C_{3v}$ ground state acceptor character and to assume a $T_{d}$ symmetry. Moreover, the charge density of the impurity level in the fault plane is higher than in the hexagonal bulk-like region, which further explains the high energetic instability.

The situation is entirely different in the case of donors (P, Sb) (see Figure~\ref{fig:charge_density}b). Here, the presence of the stacking fault does not affect the local symmetry of the ground state wave function, as donors can form four equivalent bonds with Si and have an extra electron loosely bonded apart from the host Si phase. Indeed, looking at the three plots shown in Figure~\ref{fig:charge_density}, the symmetry of the wave function does not change with respect to that observed in $n$-type doped bulk hex-Si. This behavior explains the negligible DSE observed for donors close to the fault planes. Indeed, in this case, the I$_3$-BSF is not a barrier for impurities because the steric effect is less pronounced (see Figure~\ref{fig:energy_mismatch_si}) and keeps the donor wave function symmetry.

Finally, it is worth analyzing the case of C (see Figure~\ref{fig:charge_density}c), which shows a tendency towards segregation into the fault. Looking at its charge density distribution, we can observe that when the C atom is close to the defected region, its wave function also spreads over the hexagonal region, thus inducing ---as already observed for B---  negative values of DSE.

\paragraph{Hex/cub-Si interface.} We also examine the structural characteristics and segregation energies of the dopants at the abrupt hex/cub Si interface to validate the results obtained and the DSE trends and compare them with the I$_3$-type BSF system. The analysis employs the supercell depicted in Figure~\ref{fig:geometry_of_the_simulation_sf}b, where the red-shaded region corresponds to the hex-type phase while the unshaded region corresponds to the cub-type crystal.

\begin{figure*}
    \centering
    \includegraphics[width=0.35\textwidth]{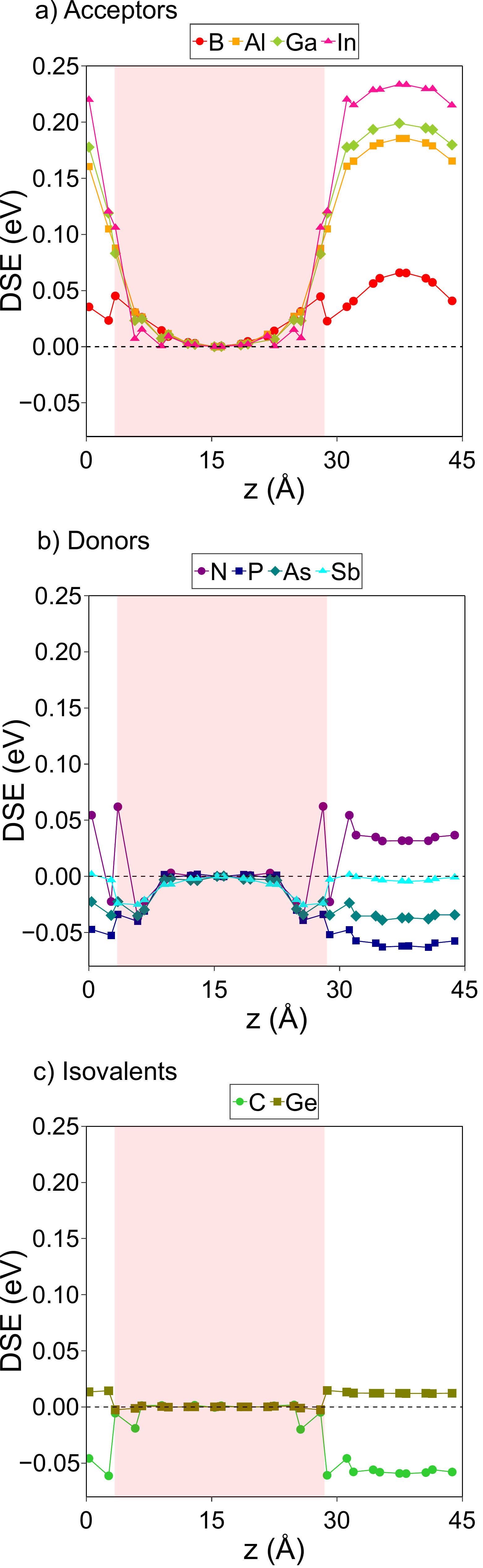}
    \caption{Segregation energy of neutral dopants at the hexagonal-diamond/cubic-diamond Si interface: (a) acceptors, (b) donors, and (c) group IV. The red region corresponds to the hexagonal-diamond phase. }
    \label{fig:energy_inter_si}
\end{figure*}

The optimized interatomic distances on both sides of the interface agree with those calculated in the fault system and those reported in other works for the corresponding bulk materials~\cite{Amato2019}. DSE calculated for this interface are reported in Figure~\ref{fig:energy_inter_si} and calculated using Eq.~\ref{eq:segregation_energy}, setting the energy reference position at the center of the hex-Si region. The results are consistent with those observed for the I$_3$-BSF system. Acceptors (Figure~\ref{fig:energy_inter_si}a) exhibit once again a significant DSE (150-250~meV), with boron being an exception and displaying a 30-40~meV energy difference near the interface. In contrast, for donors and isovalent dopants (Figure~\ref{fig:energy_inter_si}b and Figure~\ref{fig:energy_inter_si}c), the DSE variations fall within the range of tens of meV. The abrupt hex-cub Si interface can be considered the extreme case of the local configuration observed in the I$_3$-type BSF structure. Thus, these results reinforce and validate the conclusions drawn in the previous paragraphs. Although the DSE values calculated here cannot be directly compared to those of the defective case (since the stacking fault has only a local effect), they further demonstrate that the I$_3$-type BSF serves as a repulsive barrier only for acceptors. At the same time, other dopants may interact or even segregate into it.

\section{Conclusions}
The dopant interaction with I$_3$-type BSFs in hexagonal-diamond Si was investigated using first-principles ground-state DFT simulations. The structural and energetic analysis revealed that the steric effects, the dopant charge state, and the symmetry of the impurity level wave function strongly influence the stability close to the I$_3$-BSF. Neutral acceptors, except for boron, remain away from the stacking fault planes, exhibiting segregation energies that are always large and positive. This preference is attributed to the anisotropy of the hexagonal phase, which promotes a greater stress relief and induces $C_{3v}$ symmetry for the dopant level wave functions. In contrast, the segregation energy variation of neutral donors moving toward the stacking fault is almost zero and can become negative in its proximity. This is essentially due to the fact that donors have negligible steric effects, and the extra electron occupying an antibonding state has a minimum influence on bond conformation~\cite{Amato2019}.
As a consequence, the symmetry of the donor wave function is always kept when moving from the hexagonal bulk-like region to the extended plane defect. Similarly, isovalent dopants exhibit negligible or even negative segregation energies, as seen in the case of C. Furthermore, we demonstrated that negative and positive ionization of $p$-type and $n$-type dopants reduces their segregation energy values. This suggests that the dopant charge state plays a key role in the interaction with the fault planes. Finally, our analysis shows a strong segregation of neutral acceptors from the cubic to the hexagonal phase in the extreme case of an abrupt cub/hex-Si interface. Consequently, higher concentrations of extrinsic acceptors should be detected at the thermodynamic equilibrium on the hexagonal side of the interface.

These findings have important implications for the design and performance of hex-Si-based devices, as they suggest that the electrical activation of acceptors may alter their distribution within the crystal due to the thermodynamic driving force moving them away from I$_3$-BSFs. On the other hand, we expect that neutral and charged donor concentrations should be more prominent at the I$_3$-BSF as these dopants have lower energy at the stacking fault. These results are relevant for local probe chemical investigations to better characterize dopant segregation and activation phenomena in hex-Si. Ultimately, our work contributes to the growing toolbox for defect-dopant engineering in group IV materials and underscores its relevance for the development of high-performance and application-tailored semiconductor technologies.

\section*{Supporting Information}
DFT calculations of unrelaxed dopant segregation energies, charge density maps in undoped systems, Mulliken charge analysis, and optimized structure files (.xyz format).

\section*{Acknowledgement}
The authors acknowledge the ANR AMPHORE project (ANR-21-CE09-0007) and the ANR TULIP (ANR-24-CE09-5076). P. M. acknowledges the French Embassy in Kenya for its financial support and internship funding. Part of the high-performance computing resources for this project were granted by the Institut du développement et des ressources en informatique scientifique (IDRIS) under the allocations AD010914974 and AD010915077 via GENCI (Grand Equipement National de Calcul Intensif).

\bibliographystyle{unsrt}
\bibliography{manuscript}
\newpage

\newpage

\begin{figure*}[t]
	\includegraphics[width=\textwidth]{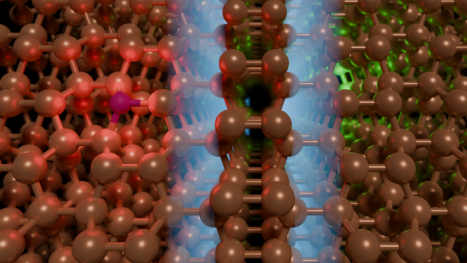}
	\caption*{TOC graphic.}
\end{figure*}

%% classes. This should be given within the "acknowledgement"

% \begin{suppinfo}

% A listing of the contents of each file supplied as Supporting Information
% should be included. For instructions on what should be included in the
% Supporting Information as well as how to prepare this material for
% publications, refer to the journal's Instructions for Authors.

% The following files are available free of charge.
% \begin{itemize}
%   \item Filename: brief description
%   \item Filename: brief description
% \end{itemize}

% \end{suppinfo}

%\bibliography{achemso-demo}

\end{document}